\documentclass[11pt,twoside]{article}

\usepackage{asp2014}

\aspSuppressVolSlug
\resetcounters

\begin{document}

\title{Decades of Transformation: Evolution of the NASA Astrophysics Data System's Infrastructure}

\author{Alberto Accomazzi}
\affil{Center for Astrophysics $|$ Harvard\ \&\ Smithsonian, Cambridge, MA, USA;
\email{aaccomazzi@cfa.harvard.edu}}

\paperauthor{Alberto Accomazzi}{aaccomazzi@cfa.harvard.edu}{0000-0002-4110-3511}{Center for Astrophysics \| Harvard \& Smithsonian}{}{Cambridge}{MA}{02138}{USA}



\begin{abstract}
The NASA Astrophysics Data System (ADS) is the primary Digital Library portal for researchers in astronomy and astrophysics. Over the past 30 years, the ADS has gone from being an astronomy-focused bibliographic database to an open digital library system supporting research in space and (soon) earth sciences. This paper describes the evolution of the ADS system, its capabilities, and the technological infrastructure underpinning it. 

We give an overview of the ADS’s original architecture, constructed primarily around simple database models. This bespoke system allowed for the efficient indexing of metadata and citations, the digitization and archival of full-text articles, and the rapid development of discipline-specific capabilities running on commodity hardware. The move towards a cloud-based microservices architecture and an open-source search engine in the late 2010s marked a significant shift, bringing full-text search capabilities, a modern API, higher uptime, more reliable data retrieval, and integration of advanced visualizations and analytics.

Another crucial evolution came with the gradual and ongoing incorporation of Machine Learning and Natural Language Processing algorithms in our data pipelines. Originally used for information extraction and classification tasks, NLP and ML techniques are now being developed to improve metadata enrichment, search, notifications, and recommendations. we describe how these computational techniques are being embedded into our software infrastructure, the challenges faced, and the benefits reaped.

Finally, we conclude by describing the future prospects of ADS and its ongoing expansion, discussing the challenges of managing an interdisciplinary information system in the era of AI and Open Science, where information is abundant, technology is transformative, but their trustworthiness can be elusive.

\end{abstract}



\section{Introduction}
The ADS as it is known today consists of a literature centric system connected to a variety of research products and enhanced by a number of services. Originally, this functionality was implemented as a part of a larger networked system which was designed to allow for the discovery, retrieval, and analysis of astrophysical data \citep{Murray1992}. That larger system was abandoned in the mid-90s in favor of a distributed set of web-based archives, but the literature component (named the “ADS Abstract Service”) survived, and later thrived, on its own.

The early literature indexing system that was developed by a small team at the Smithsonian Astrophysical Observatory in the early nineties has grown to become a discovery system which is now an essential tool not only for astronomers but for an increasing number of researchers in the space sciences. Today, an expanded version of ADS is becoming the centerpiece of NASA’s open science initiatives which aim to promote transparency and access to the research efforts funded by the agency. This paper provides an historical overview of the growth of the ADS as a technological system, focusing on the inflection points in software development which informed its evolution.

\section{Early Days: 1992-2007}
Because of the requirements involved in gathering, processing, and indexing textual data, most of the software used to create and manage the ADS Abstract database and user interface had to be developed from scratch. Relational SQL-based databases were already popular in the early 1990s, but their limitations in handling text fields meant that we could not store or index the text of an average abstract in a record. Additionally, the costs associated with purchasing access licenses for the database were significant, so the decision was made to use a file-based system for storing bibliographic records using UNIX-based servers (Sun/SunOS). This also meant that we had to design a system for indexing and then retrieving the content, thus rolling our own search engine.

The advantage of using our own search system proved important when it came to enhancing the system with functionality provided by external services. In 1993, less than one year after the introduction of the Abstract Service, we implemented the capability to perform object searches via a federated query to the SIMBAD database \citep{Eichhorn1993}, developed by the CDS in Strasbourg. To our knowledge, this was the first implementation of an inter-continental networked query as part of a search service.

\subsection{The Era of Open Source Software}
As for many similar computing-intensive projects, the lack of existing software meant that the team had to develop its own. The available tools consisted of programming languages and compilers that were either part of the UNIX operating system or were available as free software, which was starting to become popular in academic and research circles. Given our needs, we focused on developing the search engine in C and the ingest pipelines in PERL, which was then considered as the most flexible language available for processing text data. Workflows and pipelines were strung together using the components provided to us by the operating system itself: UNIX pipes, interprocess communication, shell error handling, and text processing tools such as look, sort, join, uniq.

The advent of the World Wide Web signified an inflection point for the project. As early as 1993 it became clear that the Web was going to be the best framework for delivering the services that we were developing, and the availability of free software implementing the HTTP protocol and web browsers lead to the rapid adoption of web-based technologies everywhere. Up until this point access to the ADS Abstract Service required a dedicated X-windows application that used a proprietary protocol to connect to the server hosting the database. Removing this last obstacle meant that the ADS infrastructure was able to run on commodity hardware, using open source code, and open source client-server applications. The hyperlinked nature of the web also meant that the ADS user interface could be quickly developed, enhanced, and deployed without the need for any software upgrade. Figure \ref{fig1} shows the ADS abstract service query form circa 1994.
\articlefigure{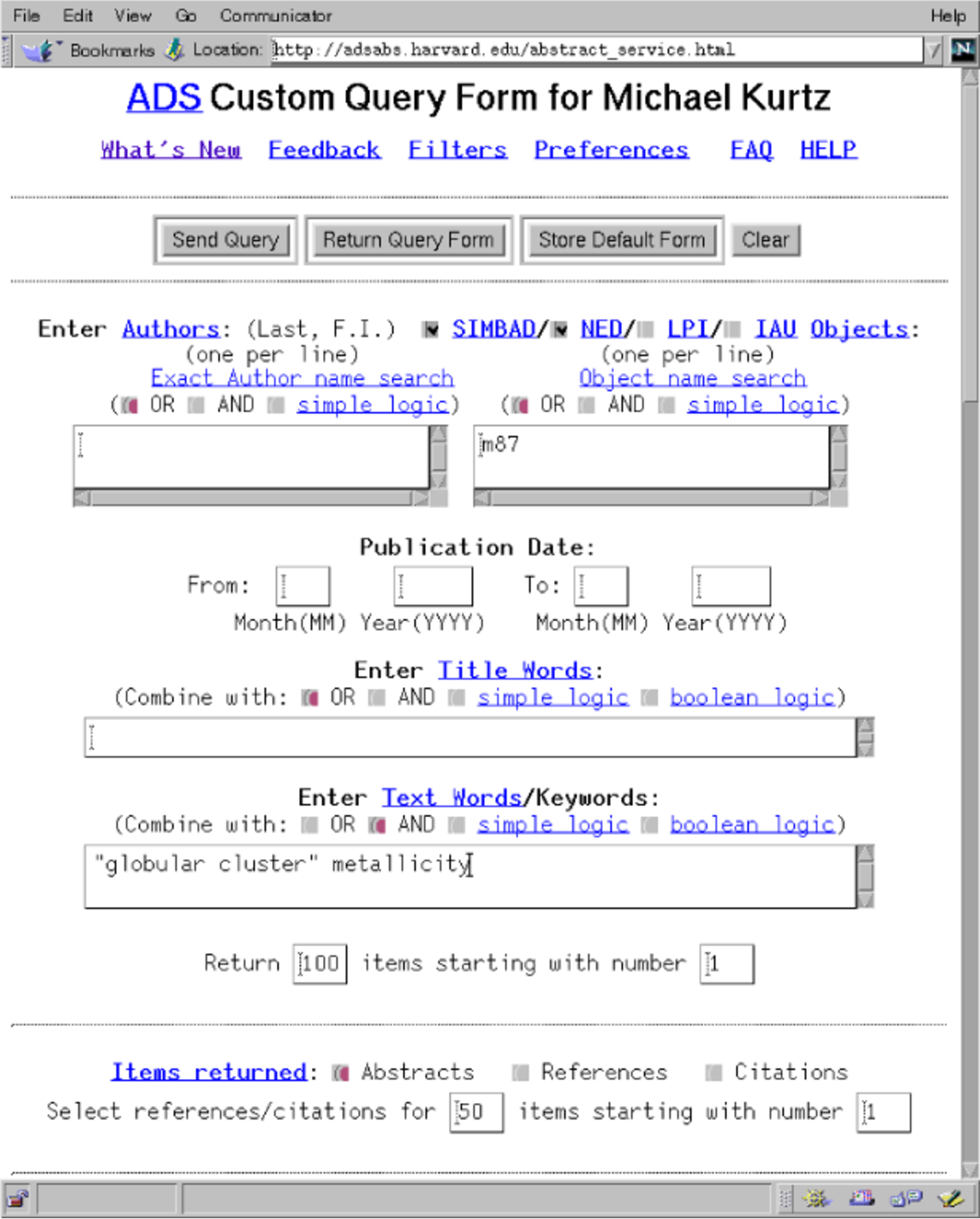}{fig1}{The beloved ADS abstract service query form, featuring over 100 search parameter settings in one single web page.}

The introduction of the GNU software suite, the Linux operating system, and the larger open source ecosystem accelerated the development and availability of tools and libraries that became useful to the project, in particular the implementation of TIFF and PNG graphic formats for our full-text document handling, the Berkeley DB system for on-disk database implementations, and POSIX threads for parallelization of tasks. Adopting the technology stack built around the open source movement meant that we had full control of our destiny, and that we could develop quickly and independently.

\subsection{The ADS Coming of Age}
Having identified the environment and platform for our system, the ADS team, which at the time was composed of 4 FTEs, was able to develop components of the web-based Abstract Service and, starting in 1995, the Article Service, which provided access to the full-text scans of the major astronomical journals \citep{Accomazzi1995}. Shortly after that, in 1996, we introduced links between bibliographic records, datasets and catalogs \citep{Accomazzi1997}. In 1997 citation data which was originally purchased from the Institute for Scientific Information by the AAS was indexed in our system \citep{Kurtz1996}, allowing users to traverse the citation graph from paper to paper, as well as to sort any search results by citation counts. 

By 1998 astronomers were hooked. A study by Kurtz (\citet{Kurtz1999}) showed that online readership through the ADS surpassed the worldwide print readership in all astronomy institutions. The availability of an increasing amount of full-text publications in the ADS and the ability for any scientist to find them and retrieve them easily and freely lead to the massive clearing of shelves from the offices of most astronomers worldwide. The embrace of the ADS platform by the community had a positive feedback effect on the evolution of the system. Publishers started hearing from their editors and referees that if their content was not in ADS, it may as well not exist, which helped the project obtain metadata and back-records of historical content in the system. 

The resounding success and adoption of the ADS in the late 1990s provided additional opportunities for the project to gain additional recognition and support from NASA, which led to an increase in staff (6 FTEs) and funds to continue the development and operations of the system. The following decade was one during which we developed system capabilities to ensure that our system was up-to-date, resilient and accessible. This led to the creation of a pipeline to automatically extract and ingest citations from the literature \citep{Demleitner1999}, a workflow to scan and digitize content on an on-going basis, and the development of a network of mirror sites hosted by collaborating institutions throughout the world \citep{Accomazzi2000}.

Additional features and services were also developed during the first decade of the 2000s. The myADS notification service was launched in 2003, providing customized digests of new publications based on the preferences of individual researchers. The original weekly updates of astronomy content from arXiv became a daily update of all the subject categories available in the preprint server. Customizations were added allowing users to save bibliographic articles in virtual private collections (“ADS libraries”), configure access to publisher content via their library’s subscriptions (based on the OpenURL protocol), and customize their search experience through a set of preferences tied to their login account.

By 2007, the ADS Abstract and Article services provided the unparalleled capabilities of any research portal in any discipline. ADS usage in 2007 shows that there were over 20,000 registered users (with login accounts), and 6,000 subscribers to the myADS notification system. Monthly usage showed that we had more than 30,000 heavy users out of an overall total of 1.2M users. Given the number of professional astronomers on earth, it was clear that the system was being used not just by the totality of researchers in the discipline, but also by people outside of astronomy as well as amateurs and the general public \citep{Henneken2009}.

The system architecture grew from one which consisted of a single abstract database to a complex set of data products and relationships, as illustrated in Figure \ref{fig2}. Greater functionality brought along greater complexity of the code, which had been growing organically for the previous 15 years. A census of our codebase conducted in 2007 showed that our search engine had grown to consist of 250K lines of C, while our indexing and application software consisted of another 250K lines of C, PERL and python 2 code. All of it custom-built and scarcely documented.  \\

\articlefigure{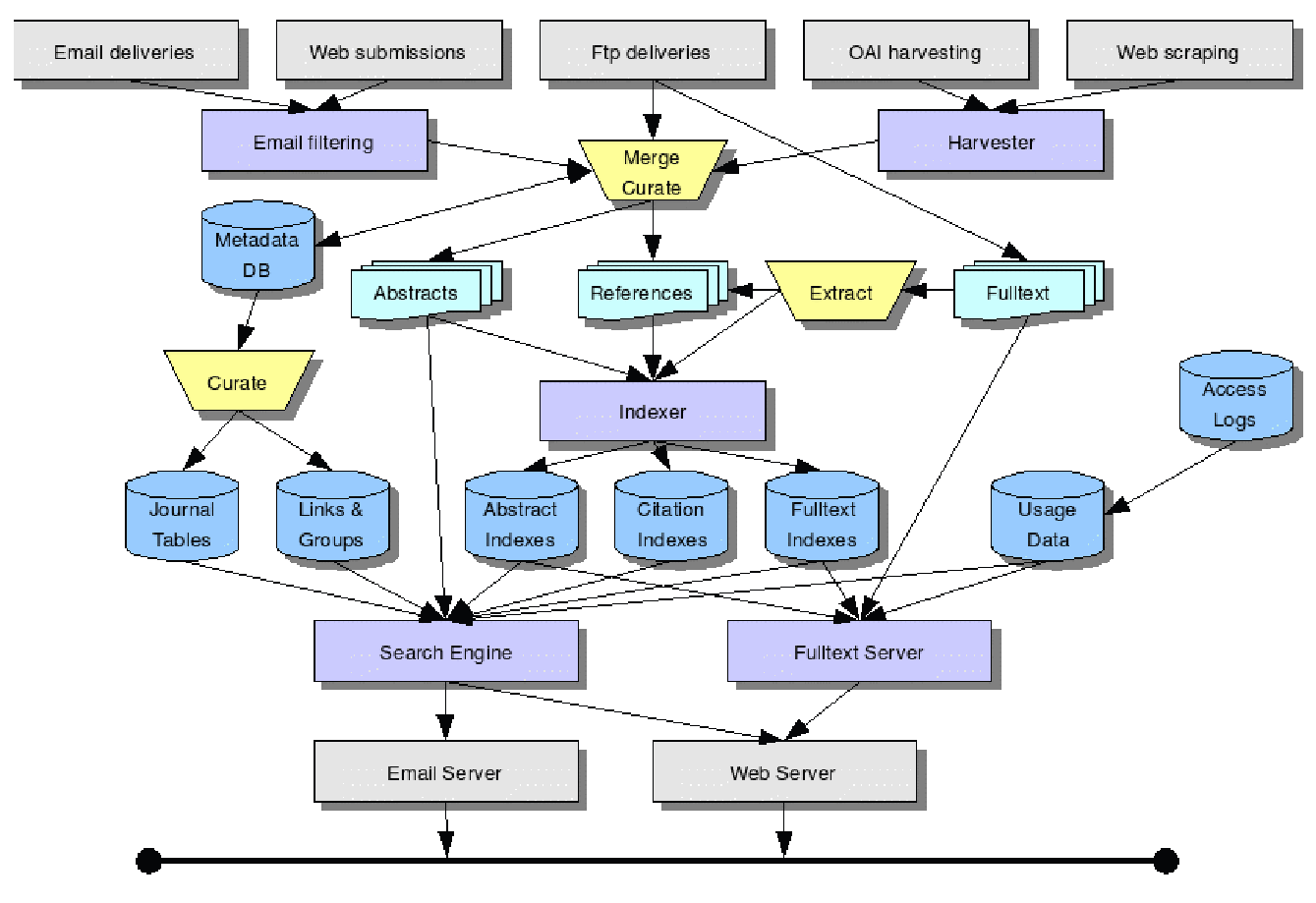}{fig2}{The ADS system architecture circa 2007. By this time the ADS system had grown into a complex set of custom-built modules and workflows.}

\section{ADS Reborn:2008-2020}
In 2007, the departure of Guenther Eichhorn, project scientist and key developer of the ADS search engine, caused a major disruption in the small team of 6 employees, and forced the project to reconsider its technical infrastructure and future development strategies. It became immediately apparent that the difficulty of maintaining and enhancing undocumented legacy code was not a wise nor sustainable path for the project, and that digital library technologies had matured to the point that open source alternatives should be considered instead. 

The task before us was not small: keep the ADS running 24/7 while rebuilding, more or less from scratch, a new system with equal or better performance and functionality. Drawing from our recent experience with system development, an additional self-imposed requirement was to develop such a system in the open, reusing as much as possible existing technologies and seeking strategic partnerships whenever appropriate. Thankfully NASA was receptive to our plea for help and supported the plan we submitted to the Astrophysics Archives review in 2008, which led to an incremental doubling in ADS staffing over the next decade.

Starting in 2009, we were able to start hiring new developers and experiment with a variety of technologies to inform our future architecture. The initial phase of this effort involved leveraging the existing search engine to document its capabilities and explore the effort required in implementing a new user interface. A pivotal shift occurred with the adoption of Apache Solr, a widely recognized open-source search platform, to manage the backend metadata indexing. This strategic change was implemented in 2013, marking a critical step in modernizing the system's architecture and setting the foundation for our search infrastructure. 

Further advancement was made in 2015 with the development and launch of a custom JavaScript application, internally codenamed “Bumblebee.” This application represented a major forward leap in terms of user interface and experience. It was during this phase that cloud computing was integrated into the system's architecture, signifying a move towards more scalable and flexible infrastructure. The cloud-based aspect of the architecture was particularly notable, with the JavaScript application utilizing a JSON API hosted in the cloud, thereby enhancing data accessibility and system responsiveness.

The subsequent phase focused on expanding the system's capabilities in order to reach feature parity with the ADS Classic system and to incorporate the evolving needs of users. One of the key enhancements was the integration of ORCID claiming, responding to the growing necessity for interoperability with academic and research author identification systems. Alongside this, considerable efforts were devoted to backend infrastructure development, ensuring the robustness and reliability required to support new functionalities and user demands.

By 2019, this comprehensive developmental journey culminated in achieving and even surpassing feature parity with the original ADS system. This milestone was not just about matching the existing functionality but also about providing an enhanced, modern platform that could support the evolving requirements of users in a more efficient, scalable, and sustainable way.  The new system featured a search engine which integrated high-performance text search with citation and usage graphs; a well-structured and well-documented JSON API; and a modern user interface featuring a number of visualizations and analytics \citep{Chyla2015}. The corresponding architecture is illustrated in Figure \ref{fig3}.
\articlefigure{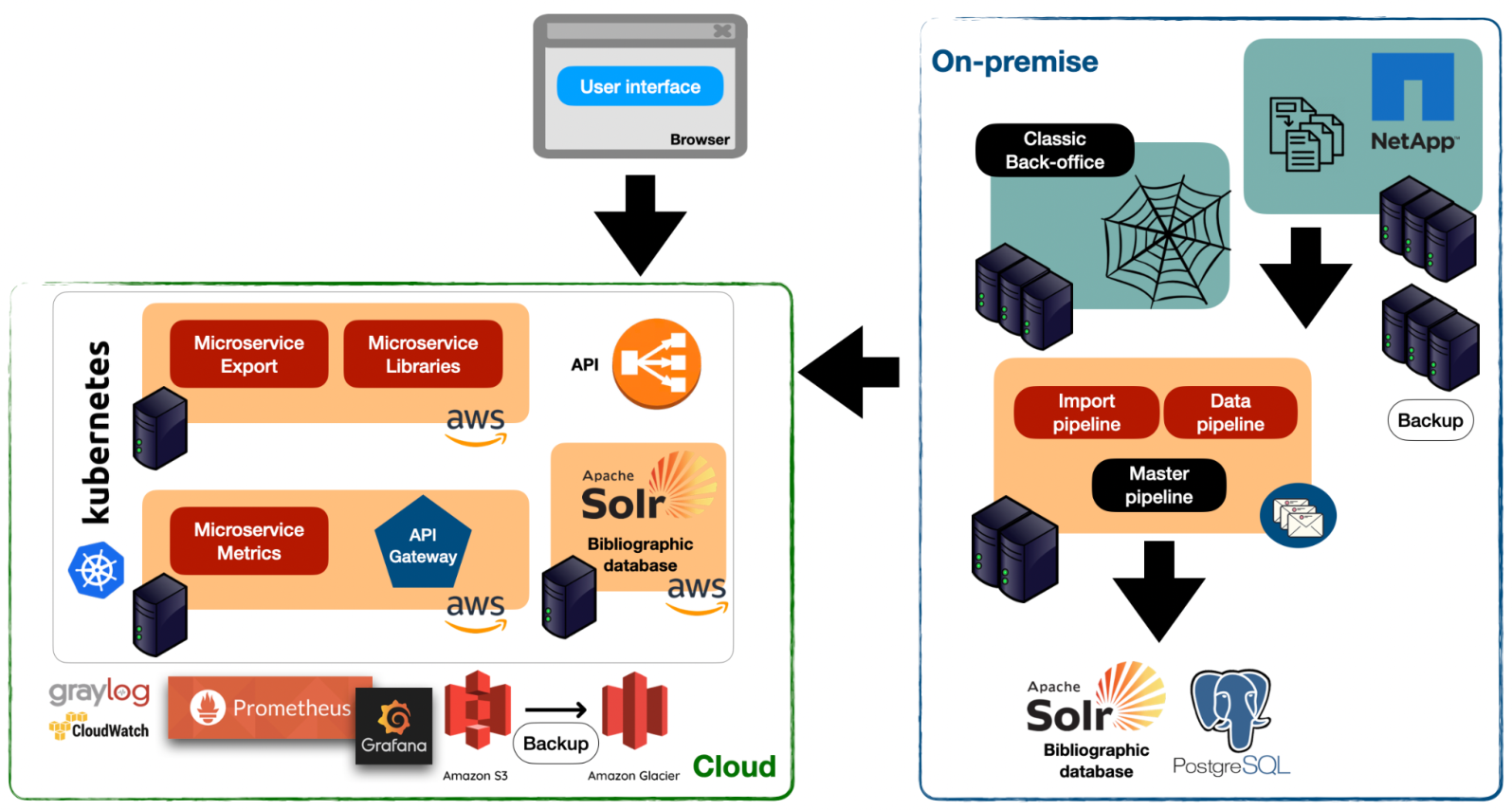}{fig3}{The current (and still evolving) ADS architecture consists of cloud-based microservices, a JSON API, and javascript user interface along with and on-premises pipelines which are still integrating some existing legacy services.}

\section{Future ADS: 2021-2030}
In 2019, the NASA Science Mission Directorate (SMD) published a white paper detailing a strategy for data and computing which included support for open science initiatives\footnote{\href{https://smd-cms.nasa.gov/wp-content/uploads/2023/05/SDMWGStrategy\_Final.pdf}{https://smd-cms.nasa.gov/wp-content/uploads/2023/05/SDMWGStrategy\_Final.pdf}}. One of the goals of the strategic plan was to provide support for open science initiatives through the creation of an interdisciplinary literature portal that could be used to understand how NASA data is used and to provide services to the research community across the disciplines funded by NASA SMD. It became immediately clear that ADS had been fulfilling these goals for the Astrophysics community, so NASA selected the ADS team to expand its mission to include all five SMD disciplines: Astrophysics, Planetary Science, Heliophysics, Earth Science, and Biological and Physical Sciences. The name of the expanded system is the NASA Science Explorer\footnote{\url{https://scixplorer.org}}, or SciX for short.

This expansion effort is transformative for the project, which will go from being a literature database focusing on a single research community to a multidisciplinary platform used by a much larger community of space, earth, and biological scientists. While much of the system infrastructure can be scaled up for additional content and use, a lot of the curation workflows and associated pipelines will need to be adapted to deal with new data providers, a new set of science data archives, and a much larger community of researchers with varied cultural backgrounds and habits. Most importantly, the ADS has become an essential research tool thanks to its role at the center of a nexus of archives and information services in astronomy. Becoming part of similar ecosystems in new disciplines will be a major challenge.

From a technical perspective, one of the things we are betting on is that we can use emerging technologies – specifically AI and machine learning – to help us perform a lot of the activities that traditionally we have been doing through human curation, for example metadata aggregation and enrichment. As an example, we want to automatically assign concepts drawn from the Unified Astronomy Thesaurus to all the records in our astronomy collection, or extract planetary feature names from planetary science papers. And part of doing this as our contribution to open science efforts means that we are not only building and delivering an AI-enhanced service, but we will also be generating and sharing the underlying training datasets and open source code that we hope will be used by all of the communities that we serve.

ADS already has a long tradition of openness, and with the expansion into SciX we will redouble our efforts to share our efforts with the larger research community. In 2021 we created and released a custom language model built on the astronomical literature called astroBERT \citep{Grezes2021}. We have contributed data sets for the 2022 and 2023 data challenges at the first and second Workshops on Information Extraction from Scholarly Papers\footnote{\url{https://ui.adsabs.harvard.edu/WIESP}}. In 2023 we have started working with the UniverseTBD collaboration\footnote{\url{https://universetbd.org}}, which recently released the first fine-tuned version of the popular LLaMA-2 model, named AstroLLaMA \citep{DungNguyen2023}. We have also participated in the creation of a Large Language Model (LLM) developed by the NASA Science Mission Directorate in partnership with IBM which will be released in 2024.

The future is uncertain but most of us will agree that AI will play a major role in the development of digital scholarship. The technologies that are most likely to revolutionize the way we interact with the scientific literature are LLMs and Knowledge Graphs (KGs). While LLM-powered chatbots have gotten the most attention from the general public, the adaptability of LLMs make them general purpose tools for a variety of Natural Language Processing tasks, such as structured information extraction, named entity recognition, and metadata enrichment. Knowledge Graphs can be similarly used to support information retrieval, semantic search, and metadata normalization. For an example of how ADS started using LLMs and KGs in its pipelines, see \citet{Shapurian2023}. We expect that this approach will be fundamental in developing SciX into a fully featured interdisciplinary system.

\section{Discussion}
Along with the promise of an exciting future, the latest AI technologies bring with them a lot of questions related to trust. Today’s LLMs are essentially black boxes, deep networks of billions and trillions of parameters trained on data of various quality, often in non-transparent ways. These systems are really too large for anyone to inspect them in any detail, but they rather seem to work in mysterious, if not magical, ways. However, as scientists, we have all been trained to reject “magic” and instead study complex systems in order to understand how they work, then modify their environment and behavior in order to control and adapt them to our needs. 

This is the task before us: use the body of knowledge generated by the scientific process to create AI technologies that advance knowledge and insight into the physical world. The ADS team and the community at large have begun investigating the use of open source LLMs for information retrieval and reasoning \citep{BlancoCuaresma2024, Ciuca2023}. While strategies based on an Retrieval Augmented Generation approach seems the most promising right now, these are still early days and there is a lot to be explored. Given the increasing number of open-source LLMs being generated, one interesting scenario for future development may be one in which there are a few open source LLMs being fine-tuned for specific tasks and domains, using custom curated datasets. Under this scenario, ADS and SciX would be the authoritative source of data used to train LLMs and build KGs used in the earth and space sciences to ensure their trustworthiness and completeness.

ADS has been a transformative service for astronomers, and it’s likely to be as transformative for a larger group of earth and space scientists in the near future. Its success couldn’t have been possible without the support of NASA and the existence of a larger ecosystem of open, interoperable information services within astronomy. The now universal support for open science initiatives gives us an opportunity to extend the ADS model to a wider set of disciplines, and the continued development of open source code and models means that there is still a bright future ahead for the scientific enterprise.

\acknowledgements 
The ADS would not exist without NASA’s continued support over the past 30 years. We are grateful to the agency for making it possible for the system to flourish and grow into its current form. The ADS team today is composed of 20 talented individuals\footnote{https://ui.adsabs.harvard.edu/about/team/}, soon to become 30 FTEs as part of our expansion. We have all benefited from the work of those who came before us and provided the vision and focus that made the ADS indispensable in its early days. There are too many names to mention, but three stand above the rest: Steve Murray, founder of the ADS and PI until his passing in 2015; Michael Kurtz, the ADS Project Scientist and visionary who is still contributing to our ongoing efforts; and Guenther Eichhorn, who managed the project until 2007 when he left the Center for Astrophysics. Today’s ADS stands on the shoulders of these giants, without which none of this would have been possible.

\bibliography{I901}  

\begin{thebibliography}{}
\expandafter\ifx\csname natexlab\endcsname\relax\def\natexlab#1{#1}\fi
\expandafter\ifx\csname url\endcsname\relax
  \def\url#1{\texttt{#1}}\fi
\expandafter\ifx\csname urlprefix\endcsname\relax\def\urlprefix{URL }\fi
\providecommand{\eprint}[2][]{\url{#2}}

\bibitem[{{Accomazzi} et~al.(1995){Accomazzi}, {Eichhorn}, {Grant}, {Murray},
  \& {Kurtz}}]{Accomazzi1995}
{Accomazzi}, A., {Eichhorn}, G., {Grant}, C.~S., {Murray}, S.~S., \& {Kurtz},
  M.~J. 1995, Vistas in Astronomy, 39, 63

\bibitem[{{Accomazzi} et~al.(1997){Accomazzi}, {Eichhorn}, {Kurtz}, {Grant}, \&
  {Murray}}]{Accomazzi1997}
{Accomazzi}, A., {Eichhorn}, G., {Kurtz}, M.~J., {Grant}, C.~S., \& {Murray},
  S.~S. 1997, in Astronomical Data Analysis Software and Systems VI, edited by
  G.~{Hunt}, \& H.~{Payne}, vol. 125 of Astronomical Society of the Pacific
  Conference Series, 357

\bibitem[{{Accomazzi} et~al.(2000){Accomazzi}, {Eichhorn}, {Kurtz}, {Grant}, \&
  {Murray}}]{Accomazzi2000}
--- 2000, \aaps, 143, 85. \eprint{astro-ph/0002105}

\bibitem[{{Blanco-Cuaresma} et~al.(2024){Blanco-Cuaresma}, {Accomazzi},
  {Kurtz}, {Henneken}, {Lockhart}, {Grezes}, {Allen}, {Shapurian}, {Grant},
  {Thompson}, {Hostetler}, {Templeton}, {Chen}, {Koch}, {Jacovich}, {Chivvis},
  {de Macedo Alves}, {Paquin}, {Batlett}, {Polimera}, \&
  {Jarmak}}]{BlancoCuaresma2024}
{Blanco-Cuaresma}, S., {Accomazzi}, A., {Kurtz}, M.~J., {Henneken}, E.,
  {Lockhart}, K.~E., {Grezes}, F., {Allen}, T., {Shapurian}, G., {Grant},
  C.~S., {Thompson}, D.~M., {Hostetler}, T.~W., {Templeton}, M.~R., {Chen}, S.,
  {Koch}, J., {Jacovich}, T., {Chivvis}, D., {de Macedo Alves}, F., {Paquin},
  J.-C., {Batlett}, J., {Polimera}, M., \& {Jarmak}, S. 2024, these proceedings

\bibitem[{{Chyla} et~al.(2015){Chyla}, {Accomazzi}, {Holachek}, {Grant},
  {Elliott}, {Henneken}, {Thompson}, {Kurtz}, {Murray}, \&
  {Sudilovsky}}]{Chyla2015}
{Chyla}, R., {Accomazzi}, A., {Holachek}, A., {Grant}, C.~S., {Elliott}, J.,
  {Henneken}, E.~A., {Thompson}, D.~M., {Kurtz}, M.~J., {Murray}, S.~S., \&
  {Sudilovsky}, V. 2015, in Astronomical Data Analysis Software an Systems XXIV
  (ADASS XXIV), edited by A.~R. {Taylor}, \& E.~{Rosolowsky}, vol. 495 of
  Astronomical Society of the Pacific Conference Series, 401.
  \eprint{1503.05881}

\bibitem[{{Ciuc{\u{a}}} \& {Ting}(2023)}]{Ciuca2023}
{Ciuc{\u{a}}}, I., \& {Ting}, Y.-S. 2023, Research Notes of the American
  Astronomical Society, 7, 193. \eprint{2304.05406}

\bibitem[{{Demleitner} et~al.(1999){Demleitner}, {Accomazzi}, {Eichhorn},
  {Grant}, {Kurtz}, \& {Murray}}]{Demleitner1999}
{Demleitner}, M., {Accomazzi}, A., {Eichhorn}, G., {Grant}, C.~S., {Kurtz},
  M.~J., \& {Murray}, S.~S. 1999, in American Astronomical Society Meeting
  Abstracts, vol. 195 of American Astronomical Society Meeting Abstracts, 82.09

\bibitem[{{Dung Nguyen} et~al.(2023){Dung Nguyen}, {Ting}, {Ciuc{\u{a}}},
  {O'Neill}, {Sun}, {Jab{\l}o{\'n}ska}, {Kruk}, {Perkowski}, {Miller}, {Li},
  {Peek}, {Iyer}, {R{\'o}{\.z}a{\'n}ski}, {Khetarpal}, {Zaman}, {Brodrick},
  {Rodr{\'\i}guez M{\'e}ndez}, {Bui}, {Goodman}, {Accomazzi}, {Naiman},
  {Cranney}, {Schawinski}, \& {UniverseTBD}}]{DungNguyen2023}
{Dung Nguyen}, T., {Ting}, Y.-S., {Ciuc{\u{a}}}, I., {O'Neill}, C., {Sun},
  Z.-C., {Jab{\l}o{\'n}ska}, M., {Kruk}, S., {Perkowski}, E., {Miller}, J.,
  {Li}, J., {Peek}, J., {Iyer}, K., {R{\'o}{\.z}a{\'n}ski}, T., {Khetarpal},
  P., {Zaman}, S., {Brodrick}, D., {Rodr{\'\i}guez M{\'e}ndez}, S.~J., {Bui},
  T., {Goodman}, A., {Accomazzi}, A., {Naiman}, J., {Cranney}, J.,
  {Schawinski}, K., \& {UniverseTBD} 2023, arXiv e-prints, arXiv:2309.06126.
  \eprint{2309.06126}

\bibitem[{{Eichhorn}(1993)}]{Eichhorn1993}
{Eichhorn}, G. 1993, JAAVSO, 22, 136

\bibitem[{{Grezes} et~al.(2021){Grezes}, {Blanco-Cuaresma}, {Accomazzi},
  {Kurtz}, {Shapurian}, {Henneken}, {Grant}, {Thompson}, {Chyla}, {McDonald},
  {Hostetler}, {Templeton}, {Lockhart}, {Martinovic}, {Chen}, {Tanner}, \&
  {Protopapas}}]{Grezes2021}
{Grezes}, F., {Blanco-Cuaresma}, S., {Accomazzi}, A., {Kurtz}, M.~J.,
  {Shapurian}, G., {Henneken}, E., {Grant}, C.~S., {Thompson}, D.~M., {Chyla},
  R., {McDonald}, S., {Hostetler}, T.~W., {Templeton}, M.~R., {Lockhart},
  K.~E., {Martinovic}, N., {Chen}, S., {Tanner}, C., \& {Protopapas}, P. 2021,
  arXiv e-prints, arXiv:2112.00590. \eprint{2112.00590}

\bibitem[{{Henneken} et~al.(2009){Henneken}, {Kurtz}, {Accomazzi}, {Grant},
  {Thompson}, {Bohlen}, \& {Murray}}]{Henneken2009}
{Henneken}, E.~A., {Kurtz}, M.~J., {Accomazzi}, A., {Grant}, C.~S., {Thompson},
  D., {Bohlen}, E., \& {Murray}, S.~S. 2009, Journal of Informetrics, 3, 1.
  \eprint{0808.0103}

\bibitem[{{Kurtz} et~al.(1999){Kurtz}, {Eichhorn}, {Accomazzi}, {Grant},
  {Demleitner}, \& {Murray}}]{Kurtz1999}
{Kurtz}, M.~J., {Eichhorn}, G., {Accomazzi}, A., {Grant}, C.~S., {Demleitner},
  M., \& {Murray}, S.~S. 1999, D-Lib Magazine, 5

\bibitem[{{Kurtz} et~al.(1996){Kurtz}, {Eichhorn}, {Accomazzi}, {Grant}, \&
  {Murray}}]{Kurtz1996}
{Kurtz}, M.~J., {Eichhorn}, G., {Accomazzi}, A., {Grant}, C.~S., \& {Murray},
  S.~S. 1996, in American Astronomical Society Meeting Abstracts, vol. 189 of
  American Astronomical Society Meeting Abstracts, 06.07

\bibitem[{{Murray} et~al.(1992){Murray}, {Brugel}, {Eichhorn}, {Farris},
  {Good}, {Kurtz}, {Nousek}, \& {Stoner}}]{Murray1992}
{Murray}, S.~S., {Brugel}, E.~W., {Eichhorn}, G., {Farris}, A., {Good}, J.~C.,
  {Kurtz}, M.~J., {Nousek}, J.~A., \& {Stoner}, J.~L. 1992, in European
  Southern Observatory Conference and Workshop Proceedings, vol.~43 of European
  Southern Observatory Conference and Workshop Proceedings, 387

\bibitem[{{Shapurian} et~al.(2023){Shapurian}, {Kurtz}, \&
  {Accomazzi}}]{Shapurian2023}
{Shapurian}, G., {Kurtz}, M.~J., \& {Accomazzi}, A. 2023, arXiv e-prints,
  arXiv:2312.08579. \eprint{2312.08579}

\end{thebibliography}


\end{document}